\documentclass[a4paper,11pt,accepted=2024-09-14]{quantumarticle}
\pdfoutput=1
\usepackage[utf8]{inputenc}
\usepackage[english]{babel}
\usepackage[T1]{fontenc}
\usepackage{amsmath}
\usepackage{hyperref}

\usepackage{tikz}
\usepackage{lipsum}

\usepackage{graphicx}
\usepackage{subfigure}
\usepackage{verbatim}
\usepackage{dcolumn}
\usepackage{bm}
\usepackage{epsf}
\usepackage{color}
\usepackage[toc]{appendix}
\usepackage{stmaryrd}
\usepackage{hhline}
\usepackage{amssymb}
\usepackage[normalem]{ulem}
\expandafter\let\csname equation*\endcsname\relax
\expandafter\let\csname endequation*\endcsname\relax
\usepackage{xstring}

\newcommand\footnoteref[1]{\protected@xdef\@thefnmark{\ref{#1}}\@footnotemark}
\makeatother

\newcommand\myeqref[1]{
	Eq. (\textup{\ref{#1}})
}

\newcommand{\trace}[1]{\mathrm{tr}\left\{#1\right\}}
\newcommand{\ptr}[2]{\mathrm{tr_{#1}}\left\{#2\right\}}

\newcommand{\bra}[1]    {\langle #1|}
\newcommand{\ket}[1]    {| #1 \rangle}

\newcommand\cF{{\mathcal F}}
\newcommand\hocom[1]{}

\newcommand{\ba}{\begin{eqnarray}}
	\newcommand{\ea}{\end{eqnarray}}
\newcommand{\bmath}{\begin{mathletters}}
	\newcommand{\emath}{\end{mathletters}}
\newcommand{\ban}{\begin{eqnarray*}}
	\newcommand{\ean}{\end{eqnarray*}}

\DeclareRobustCommand{\rchi}{{\mathpalette\irchi\relax}}
\newcommand{\irchi}[2]{\raisebox{\depth}{$#1\chi$}}

\newcommand{\tr}[1]{\mathrm{tr}\left\{#1\right\}}

\newcommand{\bla}{bla\\bla\\bla\\bla\\bla}
\usepackage{fmtcount}

\newcommand{\mc}[1]{\mathcal{#1}}

\newcommand{\draftmode}{1}    
\newcommand{\notetoself}[1]{\ifnum \draftmode=1 {\color[rgb]{0,0,0.8} [#1]} \fi}  
\newcommand{\cuttext}[1]{\ifnum \draftmode=1 {\color[rgb]{0,0.5,0} [#1]} \fi}  
\newcommand{\warntext}[1]{\ifnum \draftmode=1 {\color[rgb]{0.9,0.6,0} #1} \else {#1} \color{black} \fi}
\newcommand{\aref}[1]{{supplemental material~\hyperref[#1]{A}}}
\newcommand{\bref}[1]{{supplemental material~\hyperref[#1]{B}}}
\newcommand{\dref}[1]{{supplemental material~\hyperref[#1]{C}}}

\begin{document}

\title{Branching States as The Emergent Structure of a Quantum Universe}
	
	\author{Akram Touil}
\email{atouil@lanl.gov}
	\affiliation{Department of Physics, University of Maryland, Baltimore County, Baltimore, MD 21250, USA}
 \affiliation{Center for Nonlinear Studies, Los Alamos National Laboratory, Los Alamos, New Mexico 87545}
  \affiliation{Theoretical Division, Los Alamos National Laboratory, Los Alamos, New Mexico 87545}
	\author{Fabio Anza}
	\affiliation{Department of Physics, University of Maryland, Baltimore County, Baltimore, MD 21250, USA}
    \affiliation{Department of Mathematics Informatics and Geoscience, University of Trieste, Via Alfonso Valerio 2, Trieste, 34127, Italy}
    \affiliation{Complexity Sciences Center and Physics and Astronomy Department, University of California at Davis, One Shields Avenue, Davis, CA 95616}

    \author{Sebastian Deffner}
    \affiliation{Department of Physics, University of Maryland, Baltimore County, Baltimore, MD 21250, USA}
    \affiliation{National Quantum Laboratory, College Park, MD 20740, USA}
    
    \author{James P. Crutchfield}
    \affiliation{Complexity Sciences Center and Physics and Astronomy Department, University of California at Davis, One Shields Avenue, Davis, CA 95616}

\begin{abstract} 
Quantum Darwinism builds on decoherence theory to explain the emergence of classical behavior in a fundamentally quantum universe. {Within this framework we prove two crucial insights about the emergence of classical phenomenology, centered around quantum discord as the measure of quantumness of correlations.} First, we show that the so-called branching structure of the joint state of the system and environment is the only one compatible with zero discord. Second, we prove that for small but nonzero discord and for good but not perfect decoherence, the structure of the globally pure state must be arbitrarily close to the branching form, with each branch exhibiting low entanglement. Our results significantly improve on previous bounds and reinforce the existing evidence that this class of branching states is the only one compatible with the emergence of classical phenomenology, as described by Quantum Darwinism.
\end{abstract}

\maketitle

Why does the world appear classical? Despite striking successes in describing our quantum universe, understanding the quantum-to-classical transition is still an enigma. The core issue originates in understanding the emergence of macroscopic, predominantly classical, behavior from the specificity of microscopic quantum dynamics. More than a century after its development, quantum mechanics now offers a plethora of techniques to exploit when probing the classical limit: $\hbar \to 0$ approaches (saddle point approximations in path-integrals and WKB theory), coherent states, and high-temperature thermal states are a few that come to mind.

More recently, the rise of quantum information theory (QI), with its improved understanding of exquisitely-quantum resources, including entanglement and coherence, brought new concepts and powerful technical tools into this mix. Within the realm of equilibrium systems, notable techniques that clarify such micro-to-macro connections are found in the tools of quantum typicality \cite{Gemmer,Popescu,Deffner2015NP}, large-deviations theory \cite{Touchette}, and, more generally, concentration of measure phenomena \cite{Ledoux}. Indeed, the problem of understanding the emergence of thermal equilibrium in a quantum system exhibits similarities to the issue of understanding the emergence of classicality. Both involve a large number of degrees of freedom and macroscopic systems are, to a certain degree, both classical and thermal \cite{Le}. However, the perspectives target different questions: a classical limit does not require  equilibrium, but it can support it. 

Unequivocally, environment-induced decoherence \cite{basis1,basis2} is now recognized as a necessary element in the complex mechanism that leads a quantum system to hide its quantum-only nature in favor of a classical phenomenology. Taking one step further, Quantum Darwinism (QD)~\cite{Zurek2000AP,Zurek2003RMP,Ollivier2004PRL,Ollivier2005PRA,robin1,robin2, Zurek2009NP, strongqd1,strongqd2, touil2021eavesdropping, girolami2022quantitative,ffs1,ffs2,ffs3} then recognizes the active role played by a structured, many-body environment in explaining the emergence of classicality. In QD, the environment is understood as a communication channel, through which one learns about components of the world around us. The many-body nature of environments, therefore, becomes central, leading to the conclusion that classical information about pointer states is all that can be faithfully accessed by the different parts of an environment probing it. While the phenomenological picture of QD enjoys empirical support \cite{Unden19,Chen19,Ciampini18}, the many-body nature of environments hinders our general ability to uniquely pinpoint the structure of correlations underlying the emergence of classical phenomenology. Perhaps even more concerning is the possible non-uniqueness of such a structure, allowing for competing mechanisms to underlie the quantum-to-classical transition. This article provides clear answers to both of these challenges.

To properly frame the problem recall that entanglement, intended as a feature of a many-body system, is dual to the concept of separable states, intended as a specific structure of quantum states. That is, zero entanglement uniquely selects states that are separable. In more general terms, there is a duality between requiring that a given information-theoretic quantity has a specific value and the structure of compatible states. From this perspective, our work achieves two goals: First, we show conceptual and practical relevance of the tools of {Conditional Ensembles (CE)}, especially within the larger effort to characterize the structures of many-body quantum states compatible with a given subsystem phenomenology. Second, the intuition developed through CE leads to a rigorous proof of the following: \emph{There is one and only one quantum-state structure compatible with QD's characterization of classical behavior: singly-branching states} (as defined in Refs.~\cite{robin1,robin2}). Hereafter, for simplicity, we refer to these states as {\it branching states}. {Furthermore, it is noteworthy that the tools of conditional ensembles are well-established in the literature~\cite{pe1,pe2,pe3,pe4,pe5,pe6} and have found importance in quantum thermodynamics and, in particular, in quantum thermometry~\cite{ce1,ce2,ce3,ce4,ce5,ce6,ce7}. In the present paper, we show how CE are informative tools in quantum foundations.}

Our development is organized as follows. After a brief overview of QD, that introduces notation, we summarize CE's basic notions, as detailed in Refs.~\cite{STROCCHI1966,Ashtekar1995,Brody2001,Ashtekar1995,Gibbons1992,Heslot1985,Kibble1979,fabioGQ2,fabioGQ1,fabioGQ3,anza2024maximum} and provide a geometric description for decoherence in the pointer basis, {centered around the notion of conditional ensembles}. We proceed to show how {this ensemble picture} identifies the branching form as the unique structure for the many-body globally pure-state compatible with QD. We then state the main result and sketch the proof, which is detailed in the supplemental material. Finally, we discuss the implications of our results and provide a few forward-looking conclusions.

\paragraph{Quantum Darwinism.} QD posits a universe comprised of a quantum system $\mc{S}$ of interest and its surrounding environment $\mc{E}$, jointly described by a many-body pure state $|\psi_{\mc{SE}}\rangle$ evolving unitarily. Classicality emerges in a two-step process. $\boldsymbol{(\mathfrak{1})}$ Einselection: the environment $\mc{E}$ decoheres $\mc{S}$'s state with only the pointer states surviving \cite{basis1,basis2,zurekhalo}. $\boldsymbol{(\mathfrak{2})}$ The pure state reaches a configuration in which measurements on environmental fragments reveal almost all the classical information about $\mc{S}$. In other words, when observers make measurements on small fragments of $\mc{E}$ they reach consensus regarding the classical information inferred about $\mc{S}$'s state. 

\begin{figure}[h!]
    \includegraphics[scale=0.35]{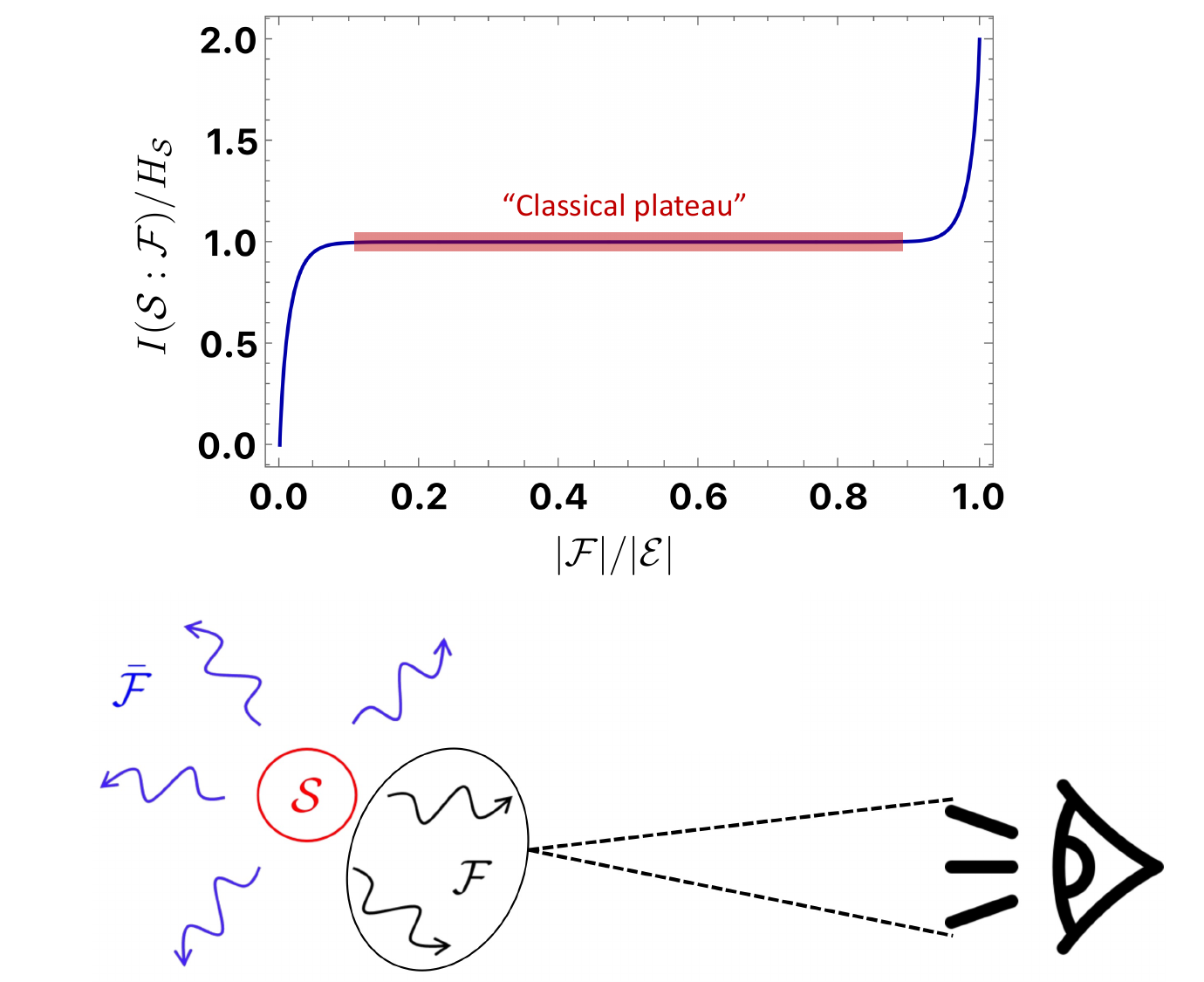}
    \caption{{An observer extracts information from a quantum system $\mathcal{S}$ embedded in a photonic environment by capturing $m$ photons $\mathcal{F}$, and inferring properties about $\mathcal{S}$. The plot shows how the mutual information $I(\mathcal{S}: \mathcal{F})$ scales as a function of the fraction $m / N \equiv|\mathcal{F}| /|\mathcal{F} \otimes \overline{\mathcal{F}}|$ of captured photons. After an initial raise, it does not matter how many photons we capture, we still only have access to approximately the same amount of information about the system. This is true unless $m \approx N$, in which case we effectively measure the entire set of photons and the quantum nature of $\mathcal{S}$ is revealed. The plateau identifies a region in which the observer can only learn classical information about the system, via pointer states.}}
    \label{illust}
\end{figure}

This implies that classical information about the pointer states proliferates across the many fragments comprising the environment. This proliferation, or redundancy, of information is the hallmark of classicality. It is witnessed by considering $\mc{E}$'s many-body nature and is recognized via an additional splitting of $\mc{E}$ into $\mc{E}=\mc{F} \otimes \overline{\mc{F}}$. That is, the environment $\mc{E}$ consists of $|\mc{F}|=m$ and $|\overline{\mc{F}}| = N-m$ fragments, respectively, where $N=|\mc{E}|$ is $\mc{E}$'s total size. Physically, $\mc{F}$ represents the environment fragments captured by the observer and $\overline{\mc{F}}$ is the rest that is inaccessible. 

{The relevant information-theoretic quantities we consider throughout the paper are the following: the quantum mutual information between system $\mc{S}$ and a fragment of the evnrionment $\mc{F}$, Holevo (accessible information), and quantum discord. The quantum mutual information is defined using the von Neumann entropy, $H_{X}=H(\rho_{X})=-\tr{\rho_{X} \log(\rho_{X})}$ as;
\begin{equation}
{I}(\mc{S}:\mc{F})=H_{\mc{S}}+H_{\mc{F}}-H_{\mc{S}\mc{F}},
\label{MUTI}
\end{equation}
where $H_{\mc{S}\mc{F}}$ is the joint entropy of system and fragment. The Holevo quantity is defined as
\begin{equation}
\chi(\mc{S}:\check{\mc{F}})= H_{\mc{S}}-H_{\mc{S}|\check{\mc{F}}},
\end{equation}
where $H_{\mc{S}|\check{\mc{F}}}$ is the entropy of the system conditioned on the outcome of measurements on $\mc{F}$. The check mark indicates the subsystem we are measuring. Finally, quantum discord is defined as~\cite{discord0},
\begin{equation}
D(\mc{S}:\check{\mc{F}})= {I}(\mc{S}:\mc{F})- \rchi(\mc{S}:\check{\mc{F}}).
\end{equation}
The quantum mutual information quantifies the total amount of bipartite correlations between system and fragment, Holevo quantifies classical correlations, while discord reflects the quantum correlations between $\mc{S}$ and $\mc{F}$.
}

{It is important to stress that the characterization of classicality is given via a range of possible values for the mutual information and discord, rather than exact values for them. This is a clear departure from previous characterizations~\cite{strongqd1,strongqd2} where the Spectrum Broadcast Structure was recognized to be compatible with the stronger assumptions of Strong Quantum Darwinism (SQD). {While conceptually clear, the assumptions of SQD entail exact values of the quantities involved and are therefore not robust to perturbations. Indeed, due to the non-linearity of these information-theoretic quantities, little can be inferred about the case in which these exact assumptions fail, even by a small amount. Additionally, SQD only characterizes the reduced state of the system and the accessible environment via the Spectrum Broadcast Structure. In this sense, SQD does not characterize the structure of the joint wavefunction compatible with local classicality. While some early work attempted to investigate this issue \cite{ffs0}, here we solve the problem in full generality.} It is noteworthy that understanding the emergence of classically, through the lenses of QD, attracted much attention in recent literature from different perspectives~\cite{ffs1,ffs2,ffs3}. However, to the best of our knowledge, none of these results establish a rigorous connection between the bipartite correlations imposed by QD and the structure of the global wave function of system and environment.}

Notably, recent advances in QD were driven by developments in quantum information theory---developments that proved crucial in tackling foundational questions. The following proposes that {conditional ensembles} should be added to the toolbox and then shows how to formulate QD in terms of {CE}. This leads directly to deriving two conditions for the nullity of discord and, eventually, to recognizing the branching form as the only state structure compatible with the emergence of classicality.

{Moreover, throughout we assume the environment decoheres the system state so that the information which the environment carries about the system refers to a unique pointer basis, i.e., single-event decoherence.}

\paragraph{{Conditional Ensembles}.} {It is well-known that a given density matrix can be written in an infinite number of different ways using different ensembles. Conditional ensembles \cite{fabioGQ1,fabioGQ2, fabioGQ3, anza2024maximum} are particular ones that answer the following question: if the environment is conditioned to be in some pure state $\ket{e_\alpha}$ by measurement, what is the resulting conditional state of the system? So, if the joint state of a system and its environment is $\ket{\psi} = \sum_{j=1}^{d_S}\sum_{\alpha=1}^{d_E}\psi_{j\alpha}\ket{a_j}\ket{e_\alpha}$} we can always use the following decomposition, determined by the state of the environment:
\begin{equation}
\ket{\psi}=\sum_{\alpha=1}^{d_E}\sqrt{p_\alpha} \ket{\chi_\alpha}\ket{e_\alpha}~.
\end{equation}
Here $p_\alpha = \bra{\psi}\mathbb{I}_S \otimes \ket{e_\alpha}\bra{e_\alpha} \ket{\psi}$ and $\ket{\chi_\alpha}=\sum_{j=1}^{d_S}\frac{1}{\sqrt{p_\alpha}}\psi_{j\alpha}\ket{a_j}$. This leads to the following ensemble of pure states of the system $\left\{p_\alpha,\ket{\chi_\alpha}\right\}_{\alpha=1}^{d_E}$ with reduced density matrix $\rho_S = \sum_\alpha p_\alpha \ket{\chi_\alpha}\bra{\chi_\alpha}$. The ensemble can be visualized geometrically as a probability measure $\mu_S$ on the projective Hilbert space of quantum states $\mathcal{P}(\mathcal{H}_S)\sim \mathbb{C}P^{d_S-1}$: $\mu_S = \sum_\alpha p_\alpha \delta_{\chi_\alpha}$, where $\delta_{\chi_\alpha}$ is the Dirac measure that identifies the pure state $\ket{\chi_\alpha}$.



In the following, we exploit these tools to build the {conditional quantum ensemble} of $\mc{S}$ and then examine QD through the lens of CE. In particular, we focus on decoherence and einselection as QD's two building blocks. The starting point---the focal point of our analysis---is the joint pure state $|\psi_{\mc{SE}}\rangle$.

\begin{figure}
    \centering
    \includegraphics[scale=0.15]{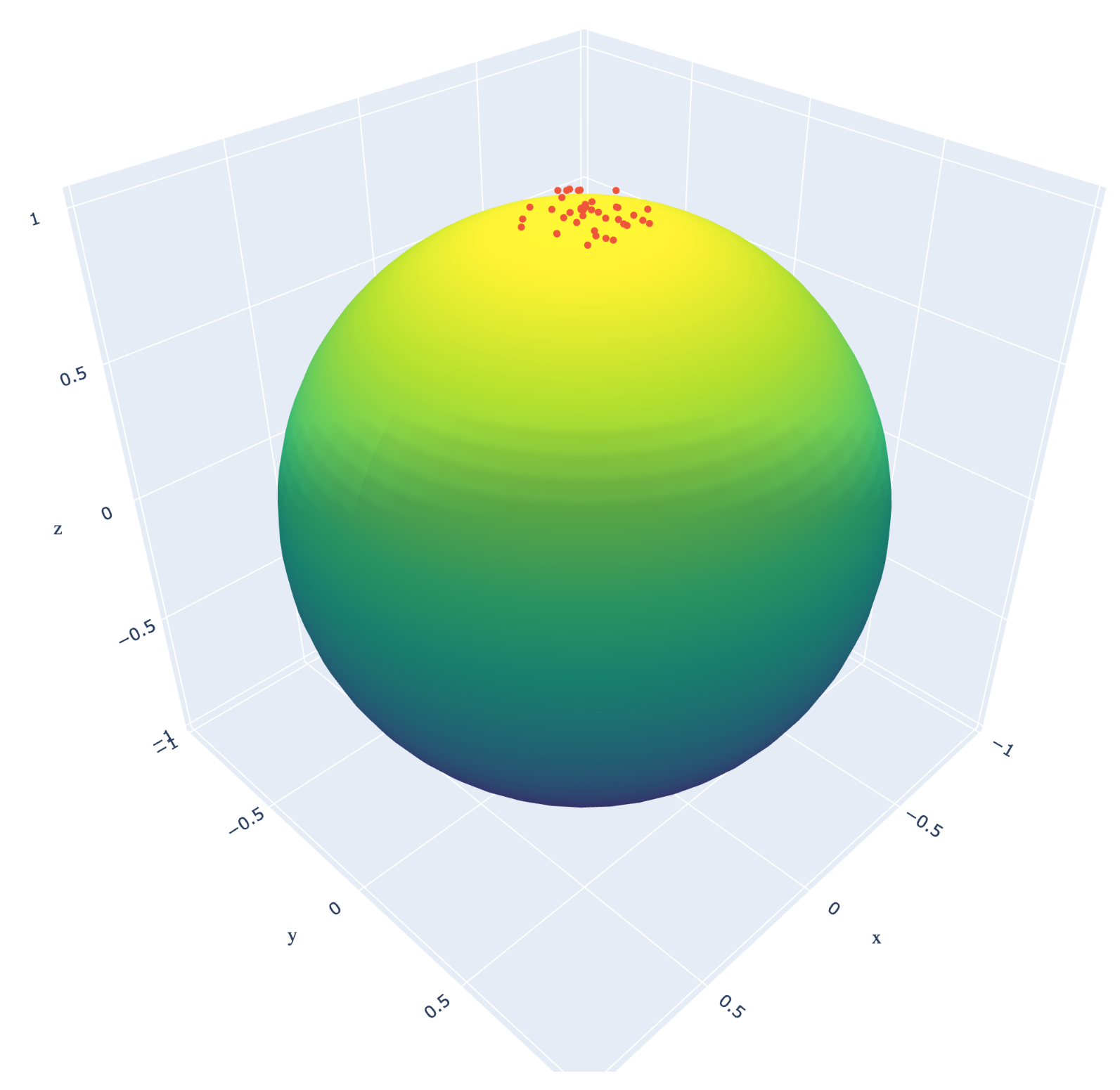}
    \caption{Decohered {conditional quantum ensemble} with pointer states $\ket{0}$ and $\ket{1}$: Red points are the $\ket{\chi_{\alpha\beta}}$ with probability mass $X_{\alpha \beta}\neq 0$ and such that the Fubini-Study distance from $\ket{0}$ is small: $D_{FS}(\ket{\chi_{\alpha \beta}},\ket{0})\leq \epsilon$. The other side of the Bloch sphere, not visible, has a qualitatively identical clustering around $\ket{1}$.}
    \label{fig:gqs}
\end{figure}

\paragraph{Decoherence.}
Any composite state $\ket{\psi_{\mc{S}\mc{E}}}$ can be written as
\begin{equation}
\begin{split}
|\psi_{\mc{SE}}\rangle\equiv |\psi_{\mc{SF\bar{F}}}\rangle&= \sum_{i, \alpha, \beta} \psi_{i\alpha\beta} |a_i\rangle |f_{\alpha}\rangle |\bar{f}_{\beta}\rangle,\\
&= \sum_{\alpha, \beta} \sqrt{X_{\alpha\beta}} |\chi_{\alpha\beta}\rangle |f_{\alpha}\rangle |\bar{f}_{\beta}\rangle,
\end{split}
\end{equation}
where $\{|a_i\rangle\}_i$, $\{f_{\alpha}\rangle\}_{\alpha}$ and $\{\bar{f}_{\beta}\rangle\}_{\beta}$ are arbitrary orthonormal bases on $\mc{H}_{\mc{S}}$, $\mc{H}_{\mc{F}}$ and $\mc{H}_{\bar{\mc{F}}}$, respectively, while $X_{\alpha\beta}=\sum_i |\psi_{i\alpha\beta}|^2$ and $\ket{\chi_{\alpha\beta}}=\sum_i \frac{\psi_{i\alpha \beta}}{\sqrt{X_{\alpha \beta}}}\ket{a_i}$. 
Following Ref.~\cite{fabioGQ1}, the {conditional quantum ensemble} of $\mc{S}$ is then $\mu_{\mc{S}}=\sum_{\alpha, \beta} X_{\alpha \beta} \delta_{\chi_{\alpha \beta}}$. This is visualized as a measure on $\mathcal{P}\left( \mc{H}_{\mc{S}}\right)$, as Fig.~\ref{fig:gqs} illustrates.

Decoherence occurs when the states $\{| \chi_{\alpha \beta}\rangle\}_{\alpha,\beta}$ cluster around pointer states. For a qubit, with pointer states $|0 \rangle$ and $|1 \rangle$, this leads to {conditional quantum ensemble}s with two clusters around antipodal points on the Bloch sphere. For a generic $\mc{S}$ of dimension $D_{\mc{S}}$, clustering around pointer states leads to
$\mu_{\mc{S}} = \sum_{n=1}^{D_{\mc{S}}} \sum_{(\alpha,\beta)\in \Omega_n} X_{\alpha \beta} \delta_{\chi_{\alpha \beta}}$,
where $\Omega_{n}$ identifies the cluster of $(\alpha, \beta)$ associated with the $n$th pointer state. The reduced density matrix of $\mc{S}$ is
\begin{equation}
\rho_{\mc{S}}= \sum^{D_{\mc{S}}}_{n,m=1} |n\rangle \langle m| \left(\sum_{(\alpha, \beta)\in (\Omega_{n} \cap \ \Omega_{m})} X_{\alpha \beta}\right).
\end{equation}
Decoherence means $\rho_{\mc{S}}$'s off-diagonal matrix elements are almost vanishing, $\sum_{\alpha,\beta \in \Omega_n \cap \Omega_m} X_{\alpha \beta}\approx 0$ for $n \neq m$. This implies that each state $\chi_{\alpha \beta}$ belongs to a unique cluster $\Omega_n$ of  nonoverlapping states, each centered around a pointer state. Before illustrating  clustering in a generic many-qubit model~\cite{touil2021eavesdropping}, we establish how the condition of zero discord results in a specific form of branching states in the universal wave function.

\paragraph{Nullity of discord.}
We explored decoherence from the perspective of {conditional quantum ensemble}s. Now, we assume decoherence happened and explore conditions for the nullity of discord to identify the organizational structure of $X_{\alpha \beta}$ and $\chi_{\alpha \beta}$ that is consistent with the vanishing of all local quantum correlations. Reference~\cite{null} already established that the nullity of discord $D(S:\check{\mc{F}})$ can be written as a condition on the form of the reduced state of $\mc{SF}$: $D(S:\check{\mc{F}})= 0 \Leftrightarrow \rho_{\mc{SF}}= \sum_{j} p_j \sigma^{j}_{\mc{S}} \otimes |f_j\rangle \langle f_j|$ {, where $\sigma^{j}_{\mc{S}}$ are density matrices of the system, conditioned on projections $|f_j\rangle \langle f_j|$ on the fragment}. We can show {the following propositions:}

\noindent{\textit{Proposition 1:} There is $\{X_{\alpha\beta}\}_{\alpha,\beta}$ such that
\begin{equation}
D(S:\check{\mc{F}})= 0 \Leftrightarrow X_{\alpha \beta} X_{\alpha^{\prime} \beta}=0, \text{for~all~} \alpha \neq \alpha^{\prime}.
\end{equation}}

\noindent{\textit{Proposition 2:} Additionally $\ X_{\alpha \beta} X_{\alpha^{\prime} \beta}=0, \forall \alpha \neq \alpha^{\prime}$, iff 
$\exists \ g(\beta)~\vert~X_{\alpha\beta}=\delta_{\alpha, g(\beta)} X_{g(\beta)\beta}$,
where $g : \mathbb{N} \to \mathbb{N}$ is a deterministic function mapping $\beta = 1,\ldots, D_{\bar{\mc{F}}}$ onto $g(\beta)=1,\ldots, D_{\mc{F}}$}\footnote{The supplemental material gives detailed proofs of both propositions.}.

These propositions establish that the only global-state structure compatible with both decoherence and exactly zero discord is given by a branching form
\begin{equation}
\begin{split}
&D(\mc{S}:\check{\mc{F}})=0 \Leftrightarrow \\
&~~|\psi_{\mc{SF}\bar{\mc{F}}}\rangle= \sum_{\beta} \sqrt{X_{g(\beta)\beta}} |\chi_{g(\beta)\beta} \rangle |f_{g(\beta)} \rangle |\bar{f}_{\beta} \rangle.
\label{discord0}
\end{split}
\end{equation}
{Moreover, we can additionally probe the above structure of states easily through the following proposition:}

\noindent{\textit{Proposition 3:}
\begin{equation}
\begin{split}
	&D(S:\check{\mc{F}})= 0  \Leftrightarrow \\ &\| \rho_{\bar{\mc{F}}}\|^{2}=\frac{1}{2} \left(\| \tilde{\rho}_{\mc{F}\bar{\mc{F}}}\|^{2}+\| \rho_{\mc{F}\bar{\mc{F}}}\|^{2}- D_{HS}(\tilde{\rho}_{\mc{F}\bar{\mc{F}}},\rho_{\mc{F}\bar{\mc{F}}})\right),
\end{split}
\end{equation}
where}
{
\begin{equation}
D_{HS}(\tilde{\rho}_{\mc{F}\bar{\mc{F}}},\rho_{\mc{F}\bar{\mc{F}}})= \trace{(\tilde{\rho}_{\mc{F}\bar{\mc{F}}}-\rho_{\mc{F}\bar{\mc{F}}})(\tilde{\rho}_{\mc{F}\bar{\mc{F}}}-\rho_{\mc{F}\bar{\mc{F}}})^{\dagger}}
\end{equation}}

{It is noteworthy that we did not explicitly make the distinction between optimal and projective measurements, since zero discord states are not perturbed by rank-1 POVMs. Therefore, in the case above, the notation ``$D(\mc{S}:\check{\mc{F}})$'' refers to optimal measurements that are also projections on $\mc{F}$. The distinction will need to be made once we extend the result to ``\textit{epsilon}'' discord.}

\paragraph{Proximity to a branching form.} The geometric framing of decoherence ({given by CE}) and discord strongly suggests that the mechanism underlying the emergence of classicality is self-organization: the elements of the support of the {conditional quantum ensemble} $\left\{ \chi_{\alpha \beta}\right\}$ spontaneously form clusters $\Omega_n$, each centered around a pointer state. In turn, this corresponds to a branching structure of the global pure state $\ket{\psi_{\mc{S}\mc{F} \overline{\mc{F}}}}$. However, as proven above, the condition holds if and only if discord is \emph{exactly} zero. Little can be said about the realistic situation in which discord is small but not exactly zero. More puzzling, due to discord's highly nonlinear character, there is no guarantee that a generic small-discord state will be in the proximity of a zero discord one. 

Entanglement has an analogous statement which pertains to the fact that a generic state with low entanglement of formation might be far from the closest separable state. While this is technically true, the geometric analysis paints a rather compelling picture, pointing towards the fact that clustering around pointer states is the only way in which classical behavior can emerge. Violation of this condition intuitively leads to violation of decoherence or large discord.

We are thus led to believe that the branching form, which is directly connected to the clustering property, might not be destroyed by a small amount of discord or small values of $\rho_{\mc{S}}$'s off-diagonal elements. This intuition, brought about by the geometric formalism, turns out to be correct. The branching structure is the only one possible when the environment is a good communication channel. This is indeed what happens when discord is small and the mutual information is near the classical plateau $I(\mc{S}:\mc{F}) \approx H_{\mc{S}}$. In point of fact, the following can be rigorously proven:

\textit{Theorem:} Given a pure state $\ket{\psi_{\mc{S}\mc{F}\overline{\mc{F}}}}$ such that $D(\mc{S},\check{\mc{F}})\leq \epsilon_D$ and $|I_{\mc{S}\mc{F}}-H_{\mc{S}}|\leq \epsilon_I$, then for all $\epsilon_D,\epsilon_I > 0$ there exists $\eta(\epsilon_D,\epsilon_I)\geq 0$ with $\eta \in \mathcal{O}(\epsilon_D,\epsilon_I)$ and a branching state $\ket{\mathrm{GHZ}}= \sum_{n=1}^{D_\mc{S}} \sqrt{y_n}\ket{n} \ket{\phi_n^{\mc{F}}}\ket{\phi_n^{\overline{\mc{F}}}}$ such that
\begin{equation}
    \left\vert \langle \psi_{\mc{S}\mc{F}\overline{\mc{F}}} \vert \mathrm{GHZ}\rangle  \right\vert^2 \geq 1-\eta(\epsilon_D,\epsilon_I)
    .
\label{mainth1}
\end{equation}
The supplemental material gives the proof.


If a state $\ket{\psi_{\mc{S}\mc{F}\overline{\mc{F}}}}$ satisfies the requirements of classicality, up to some $(\epsilon_D,\epsilon_I)$, then it will be close, in Fubini-Study distance, to a branching state: $D_{FS}(\psi_{\mc{S}\mc{F}\overline{\mc{F}}},\mathrm{GHZ}) \in \mathcal{O}(\epsilon_D,\epsilon_I)$, where $D_{FS}$ is the Fubini-Study distance. This selects the branching form as the unique structure of correlations that is compatible with classical behavior, as characterized in Quantum Darwinism. This constitutes our main result. {Mathematically, both states in Eqs.~\ref{discord0} and~\ref{mainth1} have clustering as long as we assume decoherence. The only distinction is that for Eq.~\ref{mainth1} the clustered regions are arbitrarily small and converge towards one of the pointer states. This can be seen from the fact that the form of Eq.~\ref{discord0} has degeneracies, in the sense that the index of summation ($\beta$) goes from 1 to the dimension of the complementary fragment $\overline{\mc{F}}$ while the GHZ form is exact and the summation only goes to the number of pointer states. In other words, the form of Eq.~\ref{discord0} can still lead to cases where measurements on $\mc{F}$ reveal virtually no information about $\mc{S}$.}


\begin{figure*}
    \centering
    \includegraphics[scale=0.45]{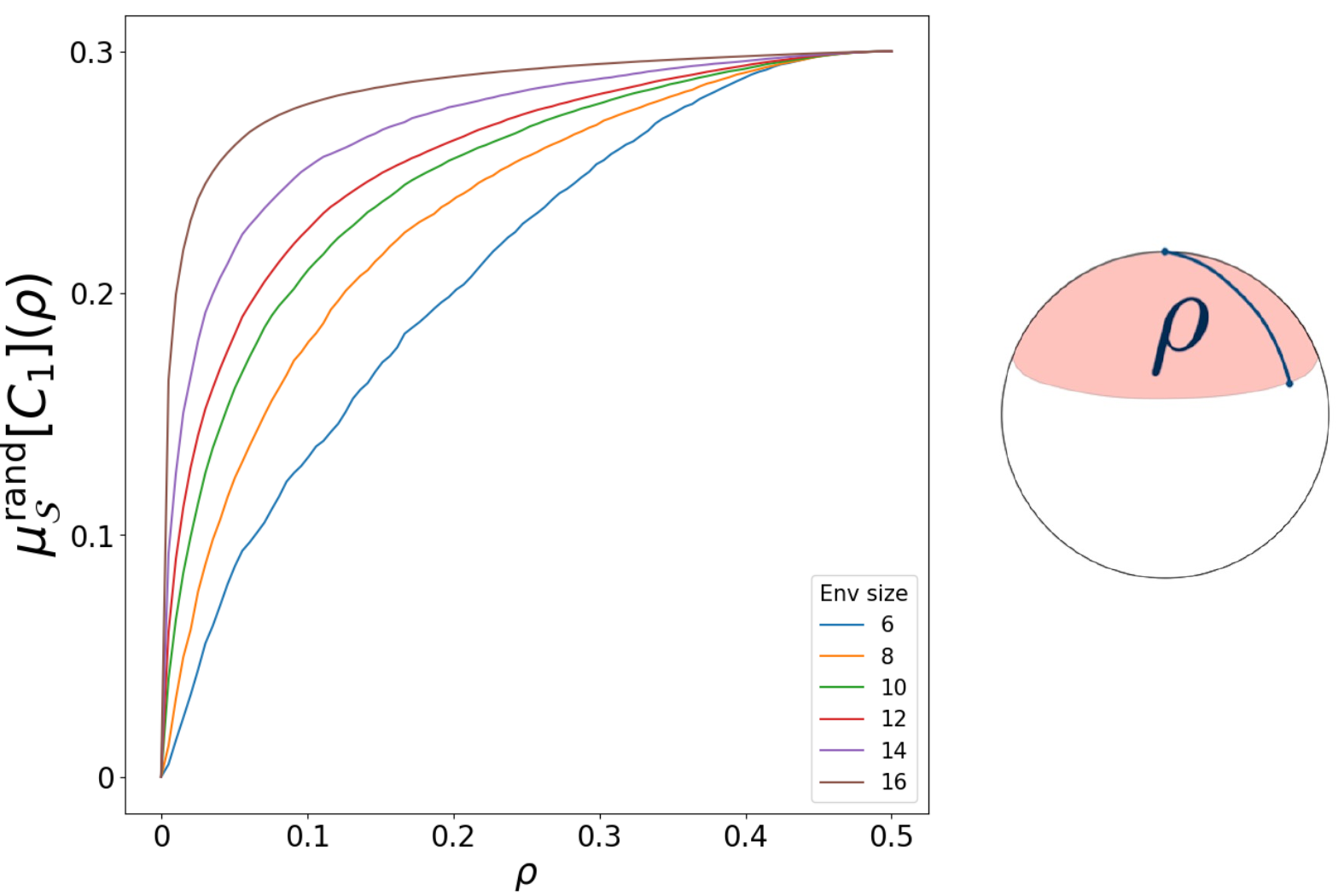}
    \caption{{Numerical experiment with dynamics generated by controlled-unitaries drawn randomly from the Haar measure, with $\mu_{\mc{S}}^{\mathrm{rand}}$ being the resulting {conditional quantum ensemble}. (Left) Probability measure $\mu^{\mathrm{rand}}_{\mc{S}}[C_1](\rho)$ of spherical caps $C_1(\rho)$ with progressively increasing Fubini-Study radius $\rho$. As the environment size increases, the curves steepen, signaling that smaller and smaller regions around $\ket{1_{\mc{S}}}$ contain increasingly large fractions of the probability mass. As $\rho \to 0.5$ we see $\mu^{\mathrm{rand}}_{\mc{S}}[C_1](\rho) \to p$, where $\sqrt{p}e^{i\phi}$ is the amplitude of $\ket{1_{\mc{S}}}$ in the system's initial state. Here, we have  $p=0.3$. (Right) Example of a spherical cap of Fubini-Study radius $\rho$, centered on $\ket{1_{\mc{S}}}$.
    The concentration of $\mu_\mc{S}^{\mathrm{rand}}$ around $\ket{1_{\mc{S}}}$ is evident. 
    While we only show concentration around $\ket{1_{\mc{S}}}$, the same holds for the other pointer state $\ket{0_{\mc{S}}}$.}}
    \label{fig:numerical_example}
\end{figure*}

\paragraph{Example: coherent information extraction.}

To illustrate the clustering property, the nullity condition in~\myeqref{zerodis}, and also provide numerical support to the validity of the main theorem, 
we use a generalized ``{\tt c-maybe}'' model. This generalizes the model of information extraction introduced and analyzed in Ref.~\cite{touil2021eavesdropping}. There, a single qubit $\mc{S}$ interacts with an environment of qubits via an imperfect {\tt c-not} gate. This is modeled with a controlled-unitary operation acting on $\mc{S}$ and the $i$th environmental qubit with ${\tt CU}_i := \ket{0_\mc{S}}\bra{0_\mc{S}} \otimes U_i^{0} + \ket{1_\mc{S}}\bra{1_\mc{S}} \otimes U_i^{1}$. In the ideal case $U_i^{0} = \mathbb{I}_{i}$ and $U_i^{1} = \sigma^x_{i}$ this is a perfect {\tt c-not} gate. The generic $U_i^{0,1}$ are meant to model a more realistic, imperfect, yet still coherent, transfer of information between a system and its environment.

To ease presentation, in the analytical section we make use of $U_i^{0} = \mathbb{I}_i$, $U_i^{1} = \sqrt{\gamma} \sigma^x_i + \sqrt{1-\gamma} \sigma_i^z$, with $\gamma \in (0,1)$, but the results are independent of this choice. This is verified in the numerical example, where we look at the model in which $U_k^{0,1}$ are randomly drawn from the Haar measure.

\paragraph{Analytical example.} If the system is initially in state $\sqrt{1-p}\ket{0_{\mc{S}}}+\sqrt{p}e^{i\phi}\ket{1_{\mc{S}}}$ and the environment is in the uniform ``all-up'' state, the dynamics leads to
\begin{equation}
\begin{split}\label{eq:cmaybe_state}
| \psi_\mathcal{SE} \rangle&= \sqrt{1-p} | 0_\mathcal{S} \rangle \bigotimes_{i=1}^{m} | 0_{i} \rangle \bigotimes_{i=m+1}^{N} | 0_{i} \rangle\\ &+ \sqrt{p}e^{i\phi} | 1_\mathcal{S} \rangle \bigotimes_{i=1}^{m} | \gamma_{i} \rangle \bigotimes_{i=m+1}^{N} | \gamma_{i} \rangle,
\end{split}
\end{equation}
with $\ket{\gamma_{i}} := U_i \ket{0_i} = \sqrt{1-\gamma} \ket{0_i} + \sqrt{\gamma}\ket{1_i}$, for all $i$. 

We can now explicitly write the system's {conditional quantum ensemble}. We adopt the following multi-index notation for the environment labels that recognizes the many-body nature of an environment made of qubits: $\alpha \to \vec{S}_\mc{F}= (s_1,\ldots,s_m)$ and $\ket{f_\alpha} \to \ket{\vec{S}_{\mc{F}}}=\otimes_{k=1}^m \ket{s_k}$, with $s_k \in \left\{ 0,1\right\}$. Analogous definitions hold for $\overline{\mc{F}}$: $\beta \to \vec{T}_{\overline{\mc{F}}}$ and  $\ket{\overline{f}_\beta} \to \ket{\vec{T}_{\overline{\mc{F}}}}$, with $t_k \in \left\{ 0,1\right\}$. This then leads to a {conditional quantum ensemble} parametrized by $X_{\alpha \beta} \to X(\vec{S}_{\mc{F}},\vec{T}_{\overline{\mc{F}}})$ and $\ket{\chi_{\alpha \beta}} \to \ket{\chi(\vec{S}_{\mc{F}},\vec{T}_{\overline{\mc{F}}})}$. For the state in Eq.~(\ref{eq:cmaybe_state}), these have the following values: $X(\vec{S}_{\mc{F}},\vec{T}_{\overline{\mc{F}}}) = 1-p + p(1-\gamma)^N$ when $\vec{S}_{\mc{F}} = \vec{T}_{\overline{\mc{F}}} = \vec{0}$ and $X(\vec{S}_{\mc{F}},\vec{T}_{\overline{\mc{F}}}) = p - p(1-\gamma)^N$ if at least one of them is not $\vec{0}$; $\ket{\chi(\vec{0}_{\mc{F}},\vec{0}_{\overline{\mc{F}}})}= \frac{\sqrt{1-p}}{\sqrt{1-p+p(1-\gamma)^N}} \ket{0_{\mc{S}}}+\frac{\sqrt{p} (1-\gamma)^{N/2}}{\sqrt{1-p+p(1-\gamma)^N}} e^{i\phi}\ket{1_{\mc{S}}}$ and $\ket{\chi(\vec{S}_{\mc{F}},\vec{T}_{\overline{\mc{F}}})}=\ket{1_{\mc{S}}}$ whenever $\vec{S}_{\mc{F}}\neq \vec{0}_\mc{F}$ or $\vec{T}_{\overline{\mc{F}}} \neq \vec{0}_{\overline{\mc{F}}}$. This leads to the following exact form of the {conditional quantum ensemble} ($\lim_{k \to \infty} \gamma^k = 0$)

\begin{align}
\mu_{\mc{S}} & = (1-p) \delta_{\ket{0_{\mc{S}}}} + p \, \delta_{\ket{1_{\mc{S}}}}
  ~.
\end{align}

This provides an analytical example of the duality between the properties listed above---decoherence, clustering, $D(\mc{S}:\check{\mc{F}})\approx 0$, and $I(\mc{S}:{\mc{F}})\approx H_\mc{S}$---and the branching form for the global wave-function $\ket{\psi_{\mc{S}\mc{F}\overline{\mc{F}}}}$. This duality holds exactly in the limit $m,N \to \infty$ or up to some small value for $m,N \gg 1$. This can be verified by direct computation of the quantities involved~\cite{touil2021eavesdropping}.

\paragraph{Numerical example.} 
To strengthen the case for the emergence phenomenology revealed in the main theorem and to provide stronger evidence that this is generic, consider the opposite case in which extracting information occurs in a random way. $U_k^{0,1}$ are now drawn randomly from the Haar measure on the unitary group. We then consider joint systems with progressively larger environment size made by $N_{\mc{E}} = 6,8,10,12,14,16$ qubits. For each, we generate the {conditional quantum ensemble} $\mu_{\mc{S}}^{\mathrm{rand}}$ by applying the random gates and then extract the cumulative distribution of the probability mass contained in progressively smaller regions around $\ket{0_{\mc{S}}}$ and $\ket{1_{\mc{S}}}$. Using a Bloch sphere representation for $\mathbb{C}P^1$, these correspond to spherical caps $C_{0/1}$ centered around the north $(0)$ and south $(1)$ poles, respectively. Geometrically, these are defined as $C_0(\rho) := \left\{\psi \in \mathbb{C}P^1 : D_{FS}(\ket{\psi},\ket{0_{\mc{S}}})\leq \rho\right\}$ and analogously for $C_1(\rho)$, centered on $\ket{1_{\mc{S}}}$.

As we increase the environment's size, the cumulative functions $\mu_{\mc{S}}^{\mathrm{rand}}[C_0](\rho)$ and $\mu_{\mc{S}}^{\mathrm{rand}}[C_1](\rho)$ become increasingly steep, proving that the support of the {conditional ensemble} becomes increasingly clustered around the pointer states. Each cluster then has probability mass provided by the initial state: $1-p$ for $\ket{0_{\mc{S}}}$ and $p$ for $\ket{1_{\mc{S}}}$. Figure~\ref{fig:numerical_example} shows the results.

In summary, in the mesoscopic and macroscopic regime $m,N\gg 1$ of the {\tt c-maybe} model, the conditions for the validity of the main theorem spontaneously emerge. And, independently, we can verify the emergence of the global-state branching form, together with its duality with the clustering property of the {conditional quantum ensemble} around the pointer states: $|0_{\mc{S}} \rangle$ and $|1_{\mc{S}} \rangle$. 

\paragraph{Concluding remarks.}
{Conditional ensembles improve} our quantum toolbox by adding new tools and concepts. Here, we exploited its application to open quantum systems, as established in Refs.~\cite{fabioGQ1,fabioGQ2}, to investigate the structures of quantum states compatible with the emergence of a classical phenomenology, as prescribed by Quantum Darwinism.

We established that classical phenomenology can emerge \emph{if and only if} the global wave-function is sufficiently close to a branching form, in which each branch has sufficiently low entanglement. The emergence of classicality is therefore understood as a self-organizing process occurring in the space of quantum states in which, in the limit of large environments, the support of the {conditional ensemble} becomes clustered around the pointer states. This has been illustrated directly, both analytically and numerically, in two realistic models of information extraction.

Our work achieves two major goals: First, it is the first use of {conditional ensembles} to investigate the emergence of classicality in an open quantum system, by directly examining the structure of the global wave-function. Second, this allowed us, for the first time, to (i) give a complete structural characterization of a global wave-function with subsystems acting classically and (ii) prove that such structure is indeed unique. 

It is still actively debated if and how classical reality is consistent with a deeply quantum universe. QD gives compelling arguments for the emergence of classicality as a consequence of decoherence and observation. Our results show that classical reality is encoded in highly structured, symmetric quantum states and that the structural properties of quantum states supporting classicality are unique. This extends the results reported in Refs.~\cite{brandao2015generic,qi2021emergent}, where it was proven that quantum many-body structures lead to locally separable states. In our case, we impose a structure of correlations compatible with quantum Darwinism, which leads to a unique structure of the global wave function, not just of the local states. As necessary in quantum Darwinism, the states we consider are epsilon-discord states as opposed to epsilon-entanglement states.

Finally, to iterate, our results apply to single-event decoherence processes, where the environment probes and learns a single basis (pointer basis), resulting in a \textit{singly-branching structure}. This analysis can be generalized by considering mixed states of the system and environment $\rho_{\mc{SE}}$. Specifically, the environment may interact with multiple systems, or the system itself could consist of many degrees of freedom that interact with the environment in various ways. In such cases, instead of using the von Neumann entropy, we would need to quantify the emergence of consensus through Shannon entropy. The exploration of whether a universal structure of states exists for these \textit{multiple-branching states} or \textit{branching trees} remains an open question in the literature. Our work suggests the possibility of a direct generalization of our theorem, which we leave for future investigation.

\acknowledgements{
Fruitful discussions with Wojciech H. Zurek, Bin Yan, Kiera Salice, Zitian Zhu, Zehang Bao, Qiujiang Guo, and Rubem Mondaini are gratefully acknowledged.

A.T. acknowledges support from the Center for Nonlinear Studies and the U.S DOE under the LDRD program at Los Alamos. F.A. acknowledges support form Templeton World Charity Foundation under grant TWCF0336 and acknowledges that this work was supported in part by the U.S. Department of Energy, Office of Science, Office of Nuclear Physics, Inqubator for Quantum Simulation (IQuS) under Award Number DOE (NP) Award DE-SC0020970. S.D. acknowledges support from the U.S. National Science Foundation under Grant No. DMR-2010127. J.P.C acknowledges support from FQXi Grant number FQXi-RFP-IPW-1902 and U.S. Army Research Laboratory and the U.S. Army Research Office under grants W911NF-21-1-0048 and W911NF-18-1-0028.}

\bibliographystyle{quantum}
\bibliography{opm}

\onecolumn\newpage

\begin{center}
\textbf{\large Supplementary material\\
Branching States as The Emergent Structure of a Quantum Universe}
\end{center}

\begin{center}
{Akram Touil, Fabio Anza, Sebastian Deffner, James P. Crutchfield}
\end{center}

\setcounter{equation}{0}
\setcounter{figure}{0}
\setcounter{table}{0}
\setcounter{page}{1}
\makeatletter
\renewcommand{\theequation}{S\arabic{equation}}
\renewcommand{\thefigure}{S\arabic{figure}}

\section{Proofs of propositions}
{\textit{Proposition 1:} We can show that there is $\{X_{\alpha\beta}\}_{\alpha,\beta}$ such that
\begin{equation}
D(S:\check{\mc{F}})= 0 \Leftrightarrow X_{\alpha \beta} X_{\alpha^{\prime} \beta}=0, \text{for~all~} \alpha \neq \alpha^{\prime}.
\label{zerodis}
\end{equation}
}

\textit{Proof of proposition 1}: ($\Leftarrow$) We generally have
\begin{equation}
	\begin{split}
		\rho_{\mc{SF}}&= \sum_{\alpha, \alpha^{\prime}, \beta} \sqrt{X_{\alpha\beta}X_{\alpha^{\prime}\beta}} |\chi_{\alpha\beta}\rangle \langle \chi_{\alpha^{\prime}\beta} | \otimes |f_{\alpha}\rangle \langle f_{\alpha^{\prime}}|,\\
		&= \sum_{\alpha \neq \alpha^{\prime}, \beta} \sqrt{X_{\alpha\beta}X_{\alpha^{\prime}\beta}} |\chi_{\alpha\beta}\rangle \langle \chi_{\alpha^{\prime}\beta} | \otimes |f_{\alpha}\rangle \langle f_{\alpha^{\prime}}|+ \sum_{\alpha, \beta} X_{\alpha\beta} |\chi_{\alpha\beta}\rangle \langle \chi_{\alpha\beta} | \otimes |f_{\alpha}\rangle \langle f_{\alpha}|.
	\end{split}
\end{equation}
Therefore, if $(\forall \alpha \neq \alpha^{\prime}); \ X_{\alpha \beta} X_{\alpha^{\prime} \beta}=0$, we get
\begin{equation}
	\rho_{\mc{SF}}= \sum_{\alpha, \beta} X_{\alpha\beta} |\chi_{\alpha\beta}\rangle \langle \chi_{\alpha\beta} | \otimes |f_{\alpha}\rangle \langle f_{\alpha}|,
\end{equation}
which implies $D(S:\check{F})= 0 $. The other implication ($\Rightarrow$) is trivial. In fact, if $D(S:\check{F})= 0 $, we have~\cite{null}
\begin{equation}
	\rho_{\mc{SF}}= \sum_{j} p_j \sigma^{j}_{\mc{S}} \otimes |f_j\rangle \langle f_j|,
\end{equation}
where $p_j \sigma^{j}_{\mc{S}}= \sum_{\beta} Y_{j, \beta} |\chi_{j\beta}\rangle \langle \chi_{j\beta}  |$. Therefore,
\begin{equation}
	\rho_{\mc{SF}}= \sum_{j, \beta} Y_{j, \beta} |\chi_{j\beta} \rangle \langle \chi_{j\beta}| \otimes | f_j \rangle \langle f_j |,
\end{equation}
hence $\left(\exists \ \{X_{j\beta}\}_{j,\beta}\right)$ such that
\begin{equation}
	\rho_{\mc{SF}}= \sum_{j, j^{\prime}, \beta} \sqrt{X_{j\beta}X_{j^{\prime}\beta}} |\chi_{j\beta}\rangle \langle \chi_{j^{\prime}\beta} | \otimes |f_{j}\rangle \langle f_{j^{\prime}}|,
\end{equation}
and $(\forall \alpha \neq \alpha^{\prime}); \ X_{\alpha \beta} X_{\alpha^{\prime} \beta}=0$. \textit{QED}.

\noindent{Proposition 2: Additionally $\ X_{\alpha \beta} X_{\alpha^{\prime} \beta}=0, \forall \alpha \neq \alpha^{\prime}$, iff 
$\exists \ g(\beta)~\vert~X_{\alpha\beta}=\delta_{\alpha, g(\beta)} X_{g(\beta)\beta}$,
where $g : \mathbb{N} \to \mathbb{N}$ is a deterministic function mapping $\beta = 1,\ldots, D_{\bar{\mc{F}}}$ onto $g(\beta)=1,\ldots, D_{\mc{F}}$}

\textit{Proof of proposition 2}: The implication ($\Leftarrow$) is trivial, namely, $X_{\alpha\beta}=\delta_{\alpha, g(\beta)} X_{g(\beta)\beta}$ implies that $(\forall \ \beta); \ (\exists! \ \alpha); \ X_{\alpha \beta} \neq 0$, and $\alpha=g(\beta)$, hence $(\forall \alpha \neq \alpha^{\prime}); \ X_{\alpha \beta} X_{\alpha^{\prime} \beta}=0$. For the implication ($\Rightarrow$), we have $(\forall \alpha \neq \alpha^{\prime}); \ X_{\alpha \beta} X_{\alpha^{\prime} \beta}=0$, if we assume that there exist at least $\alpha_1$ and $\alpha_2$ such that $\alpha_1 \neq \alpha_2$, $X_{\alpha_1 \beta} =\delta_{\alpha_1, g(\beta)} X_{g(\beta)\beta}$, and $X_{\alpha_2 \beta} =\delta_{\alpha_2, g(\beta)} X_{g(\beta)\beta}$, then we get $X_{\alpha_1 \beta}X_{\alpha_2 \beta} \neq 0$ which violates the assumption we started with. \textit{QED}.

\noindent{\textit{Proposition 3:}
\begin{equation}
\begin{split}
	&D(S:\check{\mc{F}})= 0  \Leftrightarrow \\ &\| \rho_{\bar{\mc{F}}}\|^{2}=\frac{1}{2} \left(\| \tilde{\rho}_{\mc{F}\bar{\mc{F}}}\|^{2}+\| \rho_{\mc{F}\bar{\mc{F}}}\|^{2}- D_{HS}(\tilde{\rho}_{\mc{F}\bar{\mc{F}}},\rho_{\mc{F}\bar{\mc{F}}})\right),
\end{split}
\end{equation}
where}
{
\begin{equation}
D_{HS}(\tilde{\rho}_{\mc{F}\bar{\mc{F}}},\rho_{\mc{F}\bar{\mc{F}}})= \trace{(\tilde{\rho}_{\mc{F}\bar{\mc{F}}}-\rho_{\mc{F}\bar{\mc{F}}})(\tilde{\rho}_{\mc{F}\bar{\mc{F}}}-\rho_{\mc{F}\bar{\mc{F}}})^{\dagger}}
\end{equation}}
\textit{Proof of proposition 3}: We start by noting that the elements of the density matrix $\rho_{\cF\bar{\cF}}$ have the following form:
\begin{equation}
(\rho_{\cF\bar{\cF}})_{\alpha\beta; \alpha^{\prime}\beta^{\prime}}= \langle \chi_{\alpha^{\prime}\beta^{\prime}}| \chi_{\alpha\beta} \rangle \sqrt{X_{\alpha\beta}X_{\alpha^{\prime}\beta^{\prime}}}.
\end{equation}
As derived in \textit{proposition 1},
\begin{equation}
D(S:\check{\mc{F}})= 0 \Leftrightarrow (\forall \alpha \neq \alpha^{\prime}); \ X_{\alpha \beta} X_{\alpha^{\prime} \beta}=0,
\end{equation}
hence
\begin{equation}
	D(S:\check{\mc{F}})= 0 \Leftrightarrow (\forall \alpha \neq \alpha^{\prime}); \ \sum_{\beta\gamma} (\rho_{\cF\bar{\cF}})_{\alpha^{\prime}\beta; \alpha^{\prime}\gamma}(\rho_{\cF\bar{\cF}})_{\alpha\gamma; \alpha\beta}=0.
\end{equation}
Additionally,
\begin{equation}
\begin{split}
	(\forall \alpha \neq \alpha^{\prime}); & \ \sum_{\beta\gamma} (\rho_{\cF\bar{\cF}})_{\alpha^{\prime}\beta; \alpha^{\prime}\gamma}(\rho_{\cF\bar{\cF}})_{\alpha\gamma; \alpha\beta}=0\\ &\Leftrightarrow \sum_{\alpha^{\prime}\alpha}\sum_{\beta\gamma} (\rho_{\cF\bar{\cF}})_{\alpha^{\prime}\beta; \alpha^{\prime}\gamma}(\rho_{\cF\bar{\cF}})_{\alpha\gamma; \alpha\beta}=\sum_{X}\sum_{\beta\gamma} (\rho_{\cF\bar{\cF}})_{X\beta; X\gamma}(\rho_{\cF\bar{\cF}})_{X\gamma; X\beta},
\end{split}
\end{equation}
which simplifies to
\begin{equation}
D(S:\check{\mc{F}})= 0  \Leftrightarrow \sum_{\beta\gamma} \left(\sum_{\alpha^{\prime}}(\rho_{\cF\bar{\cF}})_{\alpha^{\prime}\beta; \alpha^{\prime}\gamma}\right)\left(\sum_{\alpha}(\rho_{\cF\bar{\cF}})_{\alpha\gamma; \alpha\beta}\right)=\sum_{\beta\gamma}\sum_{X}|(\rho_{\cF\bar{\cF}})_{X\gamma; X\beta}|^2,
\end{equation}
hence
\begin{equation}
	D(S:\check{\mc{F}})= 0  \Leftrightarrow \sum_{\beta\gamma} (\rho_{\bar{\cF}})_{\gamma\beta}(\rho_{\bar{\cF}})_{\beta\gamma}=\sum_{\beta\gamma}\sum_{X}|(\rho_{\cF\bar{\cF}})_{X\gamma; X\beta}|^2,
\end{equation}
which reads
\begin{equation}
	D(S:\check{\mc{F}})= 0  \Leftrightarrow ||\rho_{\bar{\cF}}||^{2}_{F}=\sum_{\beta\gamma}\sum_{X}|(\rho_{\cF\bar{\cF}})_{X\gamma; X\beta}|^2.
\end{equation}
For further simplifications, we introduce the notation: $\tilde{\rho}_{\cF\bar{\cF}}= \sum_{X} \Pi_{X} \rho_{\cF\bar{\cF}} \Pi_{X} $ (defined as a post-measurement state), where $\Pi_{X}= |X\rangle\langle X| \otimes \mathbb{I}_{\bar{\mc{F}}}$. Therefore, we get
\begin{equation}
	D(S:\check{\mc{F}})= 0  \Leftrightarrow ||\rho_{\bar{\cF}}||^{2}_{F}=\tr{\tilde{\rho}_{\cF\bar{\cF}}\rho_{\cF\bar{\cF}}},
\end{equation}
which in turn simplifies to 
\begin{equation}
	D(S:\check{\mc{F}})= 0  \Leftrightarrow \| \rho_{\bar{\mc{F}}}\|^{2}=\frac{1}{2} \left(\| \tilde{\rho}_{\mc{F}\bar{\mc{F}}}\|^{2}+\| \rho_{\mc{F}\bar{\mc{F}}}\|^{2}- D_{HS}(\tilde{\rho}_{\mc{F}\bar{\mc{F}}},\rho_{\mc{F}\bar{\mc{F}}})\right),
\end{equation}
where $D_{HS}(\tilde{\rho}_{\mc{F}\bar{\mc{F}}},\rho_{\mc{F}\bar{\mc{F}}})= \trace{(\tilde{\rho}_{\mc{F}\bar{\mc{F}}}-\rho_{\mc{F}\bar{\mc{F}}})(\tilde{\rho}_{\mc{F}\bar{\mc{F}}}-\rho_{\mc{F}\bar{\mc{F}}})^{\dagger}}$. \textit{QED}.

\section{{Main Theorem}}
The goal is to prove the following statement
\begin{equation}
		D(\mc{S}:\check{\mc{F}}) \leq \epsilon_{D} \ \ \textit{and} \ \ H_{\mc{S}}-\epsilon_{I} \leq I(\mc{S}:\mc{F}) \leq H_{\mc{S}} \Leftrightarrow 
		|\langle GHZ|\psi_{\mc{SF}\bar{\mc{F}}}\rangle|^2 = 1-\eta(\epsilon_{D},\epsilon_{I}),
		\label{maintheorem}
\end{equation}
the constant $\eta(\epsilon_{D},\epsilon_{I})$ depends on $\epsilon_{D}$ and $\epsilon_{I}$, such that $\eta(\epsilon_{D},\epsilon_{I}) \rightarrow 0$ when $\epsilon_{D} \rightarrow 0$ and $\epsilon_{I} \rightarrow 0$. Hereafter, for simplicity, we adopt the notation $\eta \equiv \eta(\epsilon_{D},\epsilon_{I})$. Equivalently, by definition of the quantum mutual information, the statement reads
\begin{equation}
	\chi(\mc{S}:\check{\mc{F}}) \geq H_{\mc{S}}-(\epsilon_{D}+\epsilon_{I}) \ \ \textit{and} \ \ H_{\mc{S}}-\epsilon_{I} \leq I(\mc{S}:\mc{F}) \leq H_{\mc{S}} \Leftrightarrow 
	|\langle GHZ|\psi_{\mc{SF}\bar{\mc{F}}}\rangle|^2 = 1-\eta,
\end{equation}
where the state $|GHZ\rangle$ refers to the elements of the set of generalized GHZ states,
\begin{equation}
\mc{G}= \left\{ |GHZ\rangle \biggr\rvert \left(\exists \ |n\rangle \in \mc{H}_{\mc{S}}, |f_n\rangle \in \mc{H}_{\mc{F}}, |\bar{f}_n\rangle \in \mc{H}_{\bar{\mc{F}}}, y_n \geq 0\right); \ |GHZ\rangle=\sum^{D_{\mc{S}}}_{n=1} \sqrt{y_n} |n\rangle |f_{n}\rangle |\bar{f}_{n}\rangle \right\}
\label{setghz}
\end{equation}
and
\begin{equation}
	|\psi_{\mc{SF}\bar{\mc{F}}}\rangle = \sum^{D_{\mc{S}}}_{n=1} \sqrt{y_n}|n\rangle |\phi^{n}_{\mc{F}\bar{\mc{F}}}\rangle.
\label{univ_psi}
\end{equation}
The second implication $(\Leftarrow)$ is trivial, therefore to prove the statement in~\myeqref{maintheorem} we only derive the first implication $(\Rightarrow)$.

{Since the environment learns about a unique pointer basis}, and from Holevo's theorem, we have
\begin{equation}
\chi(\check{\mc{S}}:\mc{F})\geq \chi(\mc{S}:\check{\mc{F}}),
\end{equation}
hence
\begin{equation}
D(\check{\mc{S}}:\mc{F})\leq D(\mc{S}:\check{\mc{F}}),
\end{equation}
which shows that epsilon discord on the fragment's side implies epsilon discord on the system's side $D(\check{\mc{S}}:\mc{F})\leq \epsilon_{D}$. It is noteworthy that the optimal measurements on $\mc{S}$ are rank-1 projections onto the pointer basis of $\mc{S}$. The epsilon discord condition, in addition to the fact that the mutual information is near the plateau, implies
\begin{equation}
 0 \leq \sum_{n}p^{\mc{S}}_n H(\rho^{\mc{F}}_n) \leq \epsilon_{I}+ \epsilon_{D},
\end{equation}
where $\rho^{\mc{F}}_n$ are the post-measurement states of the fragment $\mc{F}$, after we project $\mc{S}$ onto the pointer basis (optimal measurement).

From the universal wave function (cf.~\myeqref{univ_psi}), we have
\begin{equation}
\begin{split}
\rho_{\mc{SF}} &= \sum_{n,m} \sqrt{y_n y_m} |n\rangle \langle m| \otimes \ptr{\bar{\mc{F}}}{|\phi^{n}_{\mc{F}\bar{\mc{F}}}\rangle\langle\phi^{m}_{\mc{F}\bar{\mc{F}}}|},\\
&= \sum_{n} y_n |n\rangle \langle n| \otimes \sigma^n_{\mc{F}}+ \Delta_{\mc{SF}},\\
&= \tilde{\rho}_{\mc{SF}}+ \Delta_{\mc{SF}}.
\end{split}
\end{equation}
{In the second step above we decompose the density matrix to two cases: $m=n$ and the off-diagonal elements $m\neq n$, hence $\Delta_{\mc{SF}}$ denotes the later part}. We have
\begin{equation}
\tilde{\rho}_{\mc{SF}}= \sum_{n,k} y_n s^{n}_k |n\rangle \langle n| \otimes |k\rangle \langle k|
\end{equation}
hence
\begin{equation}
\begin{split}
D(\rho_{\mc{SF}}\| \tilde{\rho}_{\mc{SF}})&= -H_{\mc{SF}}-\tr{\rho_{\mc{SF}}\ln{\tilde{\rho_{\mc{SF}}}}},\\
&= -H_{\mc{SF}}- \sum_{n,k} \langle n,k | \rho_{\mc{SF}} |n,k \rangle \ln{y_n s^{n}_k},\\
&= -H_{\mc{SF}}- \sum_{n,k} y_n s^{n}_k \ln{y_n s^{n}_k},\\
&= \tilde{H}_{\mc{SF}}-H_{\mc{SF}},
\end{split}
\end{equation}
where $\tilde{H}_{\mc{SF}}= H(\tilde{\rho}_{\mc{SF}})=H(\sum_{n} y_n |n\rangle \langle n| \otimes \sigma^n_{\mc{F}})$, which simplifies to $\tilde{H}_{\mc{SF}}=H(y_n) + \sum_{n} y_{n} H(\sigma^n_{\mc{F}})= H_{\mc{S}}+ \sum_{n} y_{n} H(\sigma^n_{\mc{F}})$ (cf. Ref.~\cite{nielsen2002quantum}). We note that $\sum_{n} y_{n} H(\sigma^n_{\mc{F}})=\sum_{n}p^{\mc{S}}_n H(\rho^{\mc{F}}_n) \leq \epsilon_{I} + \epsilon_{D}$ (since projecting onto the pointer basis constitutes an optimal measurement on $\mc{S}$). Therefore, we get
\begin{equation}
D(\rho_{\mc{SF}}\| \tilde{\rho}_{\mc{SF}}) \leq H_{\mc{S}}-H_{\mc{SF}}+\epsilon_{I} + \epsilon_{D}.
\end{equation}

Additionally, the condition $H_{\mc{S}}-\epsilon_{I} \leq I(\mc{S}:\mc{F}) \leq H_{\mc{S}}$ is equivalent to
\begin{equation}
	-\epsilon_{I} \leq  H_{\mc{F}}- H_{\bar{\mc{F}}} \leq 0.
	\label{cdtb}
\end{equation}
From Holevo's theorem~\cite{zwolak_entropy}, we have
\begin{equation}
H_{\mc{F}} \geq \chi(\check{\mc{S}}:\mc{F}) \geq \chi(\mc{S}:\check{\mc{F}}).
\end{equation}
Now, using the fact that $(\exists \  \epsilon_{D}, \epsilon_{I}); \ \ \chi(\mc{S}:\check{\mc{F}}) \geq H_{\mc{S}}-(\epsilon_{I} + \epsilon_{D}) \ \ \textit{and} \ \ H_{\mc{S}}-\epsilon_{I} \leq I(\mc{S}:\mc{F}) \leq H_{\mc{S}}$, we get
\begin{equation}
H_{\mc{S}}- H_{\mc{F}}\leq \epsilon_{I} + \epsilon_{D}.
\end{equation}
The above implies that
\begin{equation}
H_{\mc{S}}-H_{\mc{SF}} \leq \epsilon_{I} + \epsilon_{D}.
\end{equation}
Hence we have proven that $D(\rho_{\mc{SF}}\| \tilde{\rho}_{\mc{SF}}) \leq 2 (\epsilon_{I} + \epsilon_{D})$ and $\sum_{n} y_{n} H(\sigma^n_{\mc{F}}) \leq \epsilon_{I} + \epsilon_{D}$, which shows that the states $|\phi^{n}_{\mc{F}\bar{\mc{F}}}\rangle$ are approximately product states (for each $n$), which proves that the global wave function is arbitrarily close to a GHZ state.

\end{document}